\documentclass[10pt,conference]{IEEEtran}
\IEEEoverridecommandlockouts
\usepackage{cite}
\usepackage{amsmath,amssymb,amsfonts}
\usepackage{algorithmic}
\usepackage{graphicx}
\usepackage{textcomp}
\usepackage{xcolor}
\usepackage{url}
\usepackage{hyperref}
\def\BibTeX{{\rm B\kern-.05em{\sc i\kern-.025em b}\kern-.08em
    T\kern-.1667em\lower.7ex\hbox{E}\kern-.125emX}}
\begin{document}

\title{Validation Obligations: A Novel Approach to Check Compliance between Requirements and their Formal Specification}

\author{\IEEEauthorblockN{1\textsuperscript{st} Atif Mashkoor}
\IEEEauthorblockA{\textit{Johannes Kepler University} \\
Linz, Austria \\
atif.mashkoor@jku.at}
\and
\IEEEauthorblockN{2\textsuperscript{nd} Michael Leuschel}
\IEEEauthorblockA{\textit{Heinrich Heine University} \\
D\"{u}sseldorf, Germany \\
leuschel@hhu.de}
\and
\IEEEauthorblockN{3\textsuperscript{rd} Alexander Egyed}
\IEEEauthorblockA{\textit{Johannes Kepler University} \\
Linz, Austria \\
alexander.egyed@jku.at}
}

\maketitle

\begin{abstract}
Traditionally, practitioners use formal methods predominately for one half of the quality-assurance process: verification (do we build the software right?). The other half -- validation (do we build the right software?) -- has been given comparatively little attention. While verification is the core of refinement-based formal methods, where each new refinement step must preserve all properties of its abstract model, validation is usually postponed until the latest stages of the development, when models can be automatically executed. Thus mistakes in requirements or in their interpretation are caught too late: usually at the end of the development process. In this paper, we present a novel approach to check compliance between requirements and their formal refinement-based specification during the earlier stages of development. Our proposed approach --  ``validation obligations'' -- is based on the simple idea that both verification and validation are an integral part of all refinement steps of a system.
\end{abstract}


\section{Introduction}
\label{sec:intro}
In his 1993 report on ``Formal Methods and the Certification of Critical Systems''~\cite{rushby93a}, Rushby raised concerns that due to the unprovable nature of validation as compared to verification, the former one is to a large extent neglected by the formal methods community. While great advancements have been made in the verification area, which led to several industrial-strength environments for verification, such as RODIN~\cite{abrial10b}, the validation challenge is still open~\cite{DBLP:conf/birthday/JacquotM18}. It is not like that the formal methods community is oblivious to this important activity: we have methods like requirements state machine language (RSML)~\cite{levenson99a} and software cost reduction (SCR)~\cite{heitmeyer98a}, which are designed with an active consideration to validation. However, their contribution towards validation\footnote{Many definitions of validation exist; in the context of this paper, we use the definition from \cite{EN50128}: ``...determine whether an item fits the user needs, in particular with respect to safety and quality and with emphasis on the suitability of its operation in accordance to its purpose in its intended environment.''} of systems requirements is limited in the step-wise refinement process.

One of the main impediments for the diffusion of formal methods in the software development industry is the great difficulty to validate formal models before they are implemented. This impedes the implementation of a regular quality-assurance procedure. While the verification of a formal specification emphasizes on proving its internal consistency, validation addresses another important question entirely: whether the specification is an adequate representation of the system to be developed. Validation thus includes by definition an informal assessment: that consequences derived from the specification are in agreement with observations of a part of reality or with expectations of future users. Except for few cases (mostly numerical models), validation cannot be formalized by simply checking that ``measures'' on the real objects are equal to ``predictions'' of the model. In fact, in the case of a specification of a new system, the comparison is not with reality but with ``wishes'' of stakeholders. 

Refinement-based formal methods, such as B~\cite{abrial96a} and Event-B~\cite{abrial10a}, are highly significant for the development of (safety-critical) software systems. Refinement lets engineers structure models, verify them, and generate code by gradually adding details to the higher-level model while maintaining the already established formal properties. In this context, proof obligations (POs)\footnote{A PO is a logical formula associated with the correctness claim of a given verification property.} are crucial to ensure that models of a software system under construction are consistent and each refinement step, where a coarser-grained model is refined into a finer-grained model, preserves the properties described at the coarser-grained levels. POs also help when incrementally adapting a formal model, by pinpointing critical issues after a change of the model or the specification. However, POs do not guarantee that the ``right'' system has been developed, i.e., the development captures all stakeholders' requirements. This is the task of validation -- a process which is often carried out in an unstructured manner using techniques such as animation, simulation, model checking, code review, or acceptance testing of the generated code.

The overall aim of this paper is to enhance the refinement-based formal software development process by providing a structured and systematic validation approach using ``validation obligations.'' Validation obligations (VOs) help in checking the compliance and suitability of a specification with respect to the stakeholders' requirements, and also by providing support for incremental modeling and model evolution. While classical refinement used for code generation has a {\em single\/} linear refinement chain, the VOs methodology allows {\em multiple abstract views} of the same system (for multiple stakeholders depending upon their technical skills), and also {\em multiple instantiations\/} of a generic component for testing and simulation purposes as high-level acceptance tests can also be modeled as VOs, supported by refinement concepts. 

While POs ensure the internal consistency of a formal model and guarantee correct refinement, VOs ensure that the formal model correctly captures elicited requirements, help discover missing requirements, and help stakeholders track the evolution of their requirements and uncover high-level violations after model evolution. This is built around a simple observation: step-wise formal refinement is a well-established, effective method to master the intrinsic complexity of verification, i.e., checking the consistency and well-definedess of an implementation -- or an incremental change thereof -- against its specification. As the validation of a system's implementation against its requirements is likewise challenging, it is expected that the step-wise refinement process -- breaking the task into smaller validation steps -- will enable this task to be mastered more easily and systematically. It is crucial that each refinement step preserves the validations already carried out at the higher abstraction levels.

Rest of the paper is structured as follows: Section~\ref{sec:background} provides the necessary background to understand the content of this paper. Section~\ref{sec:VO} presents the methodology used to achieve validation obligations. Section~\ref{sec:feasibility} presents results of the feasibility study conducted using the proposed approach. The paper is concluded in Section~\ref{sec:conclusion} with the proposed future work.

\section{Background}
\label{sec:background}
\paragraph{Formal specification and refinement}

Many techniques, methods, and development processes have been proposed to assist developers in constructing reliable software. We believe that state-based rigorous methods~\cite{doi:10.1002/spe.2634} are the most promising for the development of safety-critical software. They support the ``correct by construction'' paradigm through the use of mathematically-based specification languages and the notion of refinement. While in this work we use Event-B and its ecosystem as the candidate formal method, our proposed approach is generally applicable to state- and refinement-based rigorous methods and is not only limited to Event-B.

\paragraph{Validation}

Validation requires a judgment to assess whether the work is getting nearer to fulfill the needs of the ultimate customers. Validation demands that stakeholders be convinced by evidences. The techniques must allow the people who make the validation to imagine how the artifact will work out in its
environment. In traditional engineering, different sets of tools are
used to validate artifacts: blueprints, mock-ups, prototypes, or
simulations, for instance. Software engineers have fewer tools at
their disposal though.  In a typical software development, the  requirements in the beginning phase differ from the ones in the ending phase~\cite{nakatani11a}. Two reasons account for this: 
\begin{itemize}
	\item requirements are evolving: as the development progresses, users,
	consumers, and stakeholders get a better understanding of the impact
	of the future artifact on their environment. As a consequence, they
	will discover new needs or mis-adaptations;
	
	\item requirements are incomplete:  refinement requires the
	developers to make decisions on how to represent abstract structures
	or how to decompose actions.  Often, such decisions cannot be derived
	from the requirements; they introduce new ``features'' into the
	software. 
\end{itemize}

The appeal and success of Agile methodologies~\cite{martin03a} come from this issue about managing  requirements' evolution. The development process promoted by those methodologies is based on simple principles: 
\begin{enumerate}
    \item split the development into incremental {\it runs,}   
    \item each run must be short,
    \item each run must lead to a testable product, and 
    \item the last run must produce the complete implementation.
\end{enumerate}
The third step improves the odds that an anomaly or an incompleteness in the requirements will be detected as soon as possible. Since the seminal work of Boehm~\cite{boehm81a}, we know that the earlier an anomaly is detected, the cheaper it is to correct. 

The VOs approach enables a more agile-like methodology to be used for the development of safety-critical systems: it allows one to determine the effects of requirements changes and easily adapt a formal model in such a way that important requirements keep being satisfied.

As far as validation is concerned, various works have studied how linear temporal logic (LTL) properties are affected by refinement in the context of Event-B~\cite{DBLP:journals/fac/HoangSTW16} and Z~\cite{DBLP:conf/amast/DerrickS04}. These works provide foundations to optimize the checking of VOs related to model checking. The approaches based on the fusion of the goal-oriented requirements engineering method: keep all objectives satisfied (KAOS) and Event-B, such as~\cite{mashkoor10a}, are also relevant in this context as they provide the logical foundations of our work and act as its precursors. 

\paragraph{Tools and formal issues about validation}

A formal model can be executed (for validation purposes) through one of these three techniques: translation, animation, or simulation.

In principle, through refinements, by gradually lowering non-determinism, 
we move towards fully deterministic models,  which can be easily transformed into programs. This is what the translators like EB2ALL~\cite{singh13a} or Asm2C++~\cite{bonfanti20a} do. However, both translators pose strong restrictions on the models they are able to translate such as determinism (only one event enabled at each time), implementable data structures (no sets or quantified expressions), and substitutions as assignments. 

The need for executing non-deterministic formal models has been recognized since beginning and some tools have also been provided. While Brama~\cite{servat06a} is a pure animator, ProB~\cite{leuschel08a} -- a model checker -- can also be used to run animations. Animators are based on implementations of the operational semantics of the language. They can deal with some non-determinism such as choosing events to fire or choosing values for event-parameters. While they impose less constraints on specifications than translators, they still put some strong limits on the class of models that can be animated. To this end, techniques like VTA (verify-transform-animate)~\cite{mashkoor17d}, which are based on syntactic transformations of specifications can help render non-animatable specifications animatable up to certain extent. Of course, the correctness of such transformations becomes a crucial issue then.

Translator and animators may fail. Then a technique called {\it simulation}~\cite{mashkoor17c} may work. The idea is to combine translators (by generating executable implementations of formal traits), animators (by providing an engine and libraries implementing the operational semantics), and users' intelligence (by providing them with hooks where they can insert their own code to overcome non-determinism or to prevent combinatorial explosions). The simulators may not exhibit all the behaviors specified, but the judicious restriction to the ``interesting'' behaviors allows validation to take place. 

\paragraph{ProB}

ProB is a model checking software, which also provides animation capabilities. Initially, ProB was developed as a model checker for the B language. It was extended with animation facilities, which allow developers to run interactive and graphical executions of formal models. ProB now supports a variety of state-based formal methods. ProB provides formal methods users with functionalities such as deadlock checking, disproving (finding counter examples in proofs), or animation. Our proposed approach is built upon the ProB platform.

\section{Validation obligations}
\label{sec:VO}

\begin{figure}[ht]
	\begin{center}
		\includegraphics[width=0.99\linewidth]{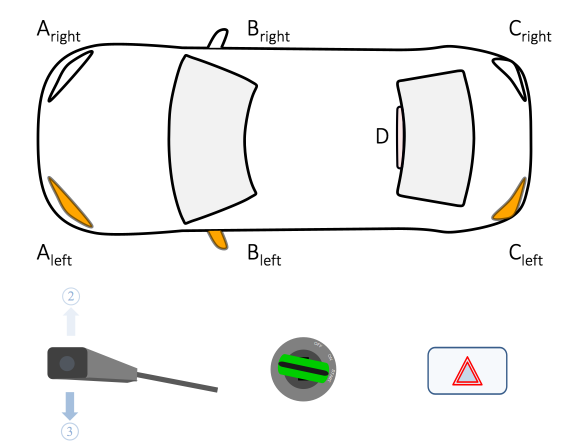}
		\caption{Graphical visualization of the state of the adaptive exterior vehicle lighting model in the prototype \label{fig:car}}
	\end{center}
\end{figure}

Our methodology to realize the VOs approach first consists in characterizing and formalizing the notion of VOs. In particular, we classify how to discharge VOs and how they are affected by refinement and by incremental model changes. Some VOs may require human intervention, e.g., to evaluate a graphical rendering of the state-space or of a particular state as shown in Fig.~\ref{fig:car}; here a stakeholder may wish to inspect that the blinking behavior of the vehicle corresponds to his/her expectations. 

\begin{figure}[ht]
	\begin{center}
		\includegraphics[width=0.99\linewidth]{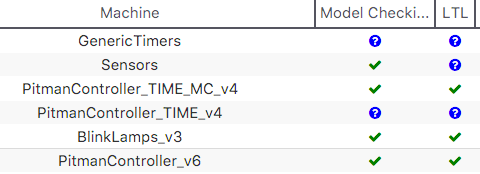}
		\caption{Project view of the prototype UI related to model checking and LTL properties checking}  
		\label{fig:projectview}
	\end{center}
\end{figure}

Other VOs, like a temporal LTL property as shown in Fig.~\ref{fig:projectview}, can be discharged automatically by model checking and when proven for an abstraction may hold for a refinement or only require a simplified additional check to be carried out. While some VOs may require further instantiations of the model, others, on the contrary, may require further abstractions.

\begin{figure}[ht]
	\begin{center}
		\includegraphics[width=0.99\linewidth]{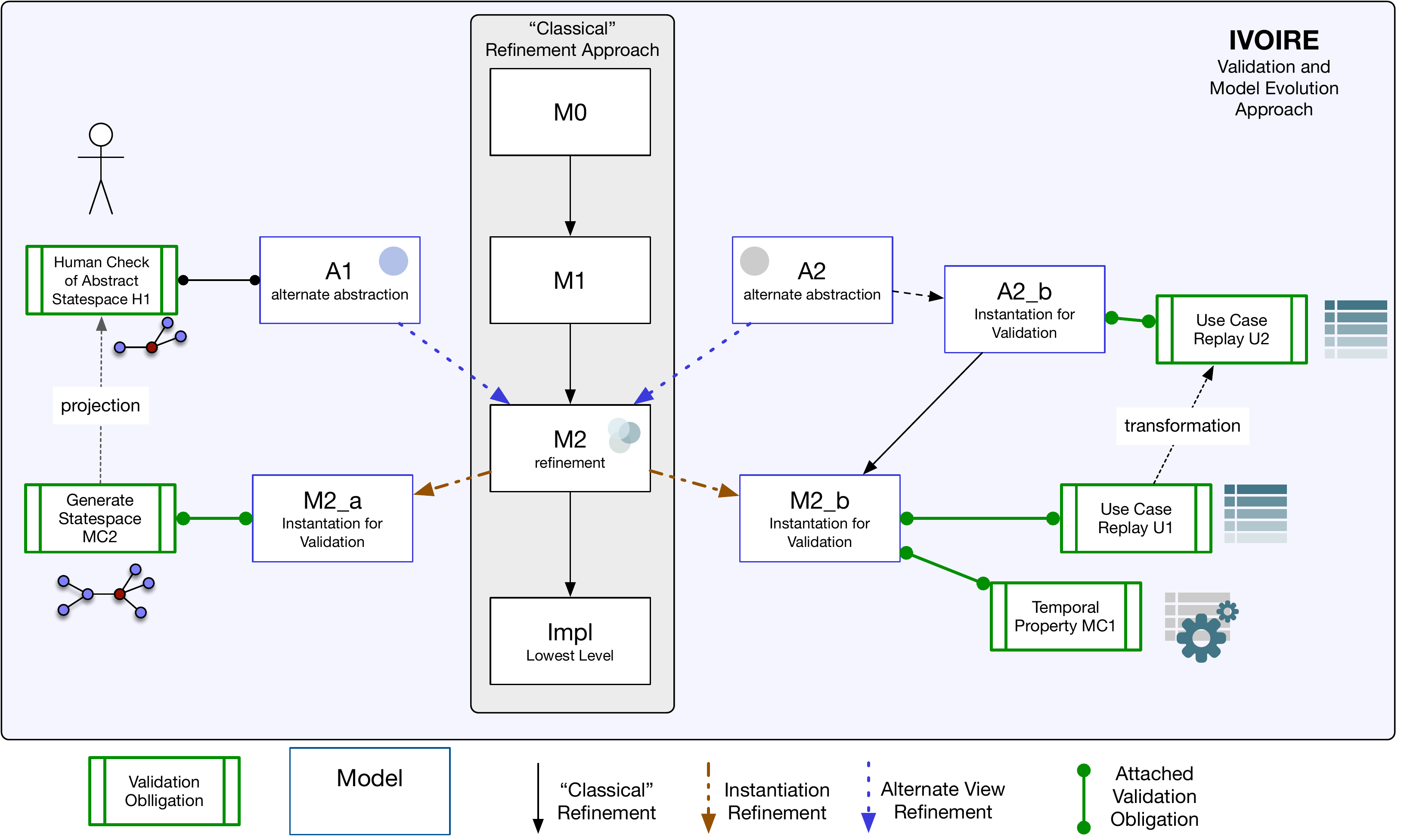}
		\caption{Nonlinear refinement sketch\label{fig:reflattice}}
	\end{center}
\end{figure}

\begin{figure}[ht]
	\begin{center}
		\includegraphics[width=0.99\linewidth]{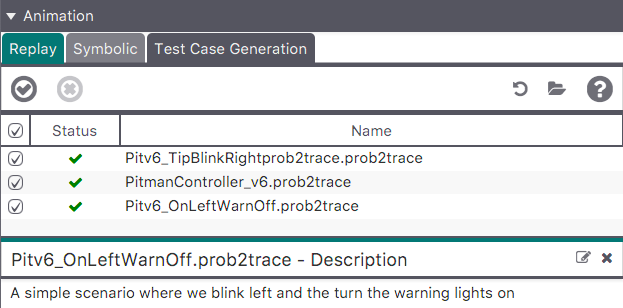}
		\caption{Stored traces/scenarios in the prototype UI \label{fig:traceview}}
	\end{center}
\end{figure}

In order to organize VOs, we hence propose to use nonlinear refinement chains. In Fig.~\ref{fig:reflattice}, M0, M1, M2, and Impl correspond to the classical linear refinement chain, where M0 is the most abstract and Impl the most concrete model. The models M2a and M2b, however, are instantiations of M2 for validation purposes, while A1 and A2 are additional abstract views, e.g., projections of the system for various stakeholders' concerns. Possible VOs here are: 
\begin{itemize}
    \item 
use cases expressed as state machines or high-level assurance tests, 
\item collections of detailed scenarios (see Fig.~\ref{fig:traceview}), 
\item concrete unit tests, 
\item temporal properties checking functional adequacy, 
\item coverage properties (MC/DC) to ensure that all parts of the model are necessary and validated, 
\item tests run on the final generated code, etc.
\end{itemize}
Similarly, an execution trace for a refinement (e.g., U1 on the right side of Fig.~\ref{fig:reflattice}) can be translated into an execution trace of an abstract view of the system.

Regarding nonlinear refinement for validation, we need to capture the fact that there can be multiple views, e.g., multiple abstractions for the same system or component. We capture multiple instantiations of generic components for validation purposes in our refinement structure. This gives rise to nonlinear refinement lattices (cf., Fig.~\ref{fig:reflattice}) rather than one linear refinement chain. Examples of nonlinear refinement are instantiation refinement or alternate view refinement as shown in Fig.~\ref{fig:reflattice}. M2 in Fig.~\ref{fig:reflattice} is, in general, not the composition of M1, A1 and A2; A1 and A2 are alternate (stakeholder) views of the system and may overlap in ways not allowed by formal decomposition.

\section{Feasibility study}
\label{sec:feasibility}
We test the feasibility of our approach on two case studies: 
\begin{enumerate}
    \item The first case study is about the development of a formal specification based on the new hybrid level 3 European train control system (ETCS) principles document~\cite{HL3}. The VOs-inspired development leads the uncovering of around 40 issues in the original requirements document, leading to changes both to the original requirements document (updated versions of \cite{HL3} are now available) as well as to our initial formal model. The visualizations of the formal model (see Fig.~\ref{hl3-prob}) were extremely useful to allow domain experts to understand and correct the issues. Managing the model evolution using the current state-of-the-art, i.e., without VOs, was very cumbersome and error prone.

\begin{figure}
 \begin{center}
  \includegraphics[width=0.99\linewidth]{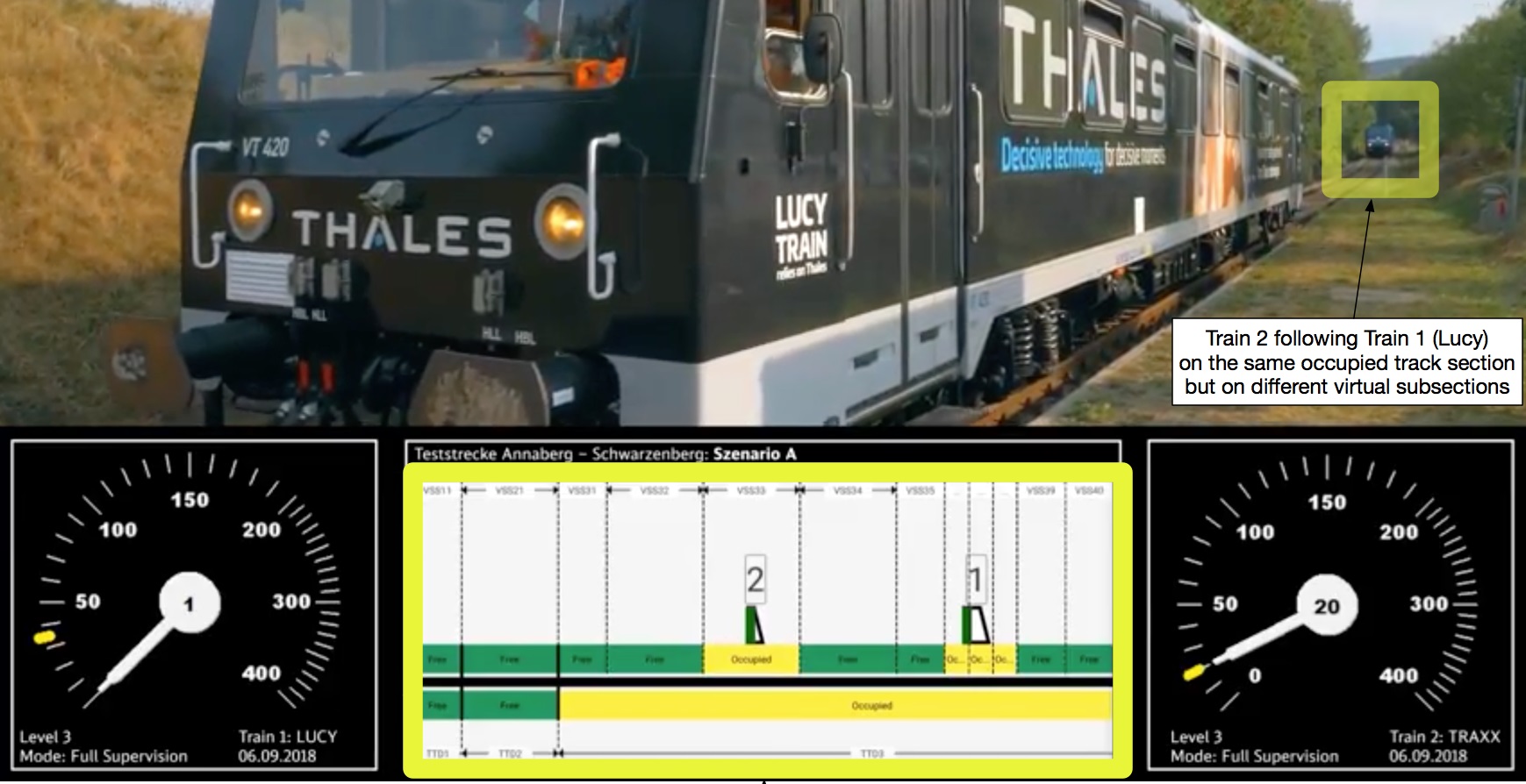}
  \end{center}
 \caption{Screen-shot from a video of Deutsche Bahn showing a formal model in action 
 \protect\url{https://www.youtube.com/watch?v=K6mS6akRmvA}}
 \label{hl3-prob}
\end{figure}

    \item The second case study is about the formal specification of an adaptive exterior light and speed control system~\cite{frank20a}. In this VOs-inspired formal development, before checking POs, we validated our models using model checking and animation with visualizations as shown in Fig.~\ref{fig:car} and Fig.~\ref{fig:projectview}. This not only helped uncover requirements errors but also helped transforming the model into an executable, interactive reference specification, which could also be examined by users without any formal background. The model checking and LTL formulas checking further helped ensuring the overall validation of the specification by breaking it into smaller steps associated with each refinement.   
\end{enumerate}

\section{Conclusion}
\label{sec:conclusion}
In this paper, we have proposed the notion of VOs to formally check the compliance between requirements and their formal specification. The presented approach is also useful in facilitating incremental modeling and tracking model evolution. We have checked the feasibility of the proposed approach on real-life industrial-strength case studies of ETCS HL 3 and the adaptive  exterior  light  and  speed  control system of a vehicle.  

Tooling is an important part of the formal methods research, both to help validate the developments and to disseminate them. To this end, we have developed a prototype (an extension of ProB), which is able to design and validate scenarios in the form of specification animation, generate and execute test cases, and perform model checking and LTL properties checking. In fact, the prototyped VOs manager keeps track of the discharged VOs, model evolution, and the information that how various VOs are linked to refinement (as shown in Fig.~\ref{fig:projectview} and Fig.~\ref{fig:traceview}). 

In the future, we intend to further develop the tool as hinted in Sec.~\ref{sec:VO}. We also intend to conduct more case studies and controlled experiments as a part of the future work.

\section*{Data availability}
The ProB tool used in this research is openly available at \url{https://prob.hhu.de}.

\section*{Acknowledgment}
The research reported in this paper has been partly funded by the Austrian Science Fund (FWF) (grant \# I 4744-N), and the LIT Secure and Correct System Lab sponsored by the province of Upper Austria.

\bibliographystyle{IEEEtran}
\bibliography{Biblio}
\end{document}